\documentclass[twocolumn,prl,10pt,amsmath,amssymb,nofootinbib,showpacs,superscriptaddress,floatfix]{revtex4-1}

\usepackage{graphicx}
\usepackage{color}
\usepackage[usenames,dvipsnames]{xcolor}
\usepackage[colorlinks=true,linkcolor=Red,citecolor=Green,linktoc=page]{hyperref}
\usepackage{multirow}
\usepackage{float}
\usepackage{flushend}
\usepackage{balance}
\usepackage[varg]{txfonts}
\usepackage{fancyhdr}

\usepackage[usenames,dvipsnames]{xcolor}
\usepackage[colorlinks=true,linkcolor=Red,citecolor=Green,linktoc=page]{hyperref}

\begin{document}

\title{Harnessing ferroelectric domains for negative capacitance}
\author{I.\,Luk'yanchuk}
\affiliation{University of Picardie, Laboratory of Condensed Matter Physics,
Amiens, 80039, France} 
\affiliation{Landau Institure for Theoretical
Physics, Moscow, Russia}
\author{ Y.\,Tikhonov}
\affiliation{Faculty of Physics, Southern Federal University, 3 Zorge str., 344090 Rostov-on-Don, Russia} 
\author{A.\,Sen\'{e}}
\affiliation{University of Picardie, Laboratory of Condensed Matter Physics,
Amiens, 80039, France} 
\author{A.\,Razumnaya}
\affiliation{University of Picardie, Laboratory of Condensed Matter Physics,
Amiens, 80039, France} 
\affiliation{Faculty of Physics, Southern Federal University, 3 Zorge str., 344090 Rostov-on-Don, Russia} 
\author{V.\,M.\,Vinokur}
\affiliation{Materials Science Division, Argonne National Laboratory, 9700
S. Cass Ave, Argonne, IL 60439, USA}

\begin{abstract}
The pressing quest for overcoming Boltzmann tyranny in low-power
nanoscale electronics revived the
thoughts of engineers of early 1930-s on the possibility of negative circuit
constants. The concept of the ferroelectric-based negative capacitance (NC) devices
triggered explosive activity in the field.
However, most of the research addressed transient NC, leaving the basic
question of the existence of the steady-state NC unresolved.
Here we demonstrate that the ferroelectric nanodot capacitor hosts a stable two-domain state realizing the static reversible NC device thus opening routes for the extensive use of the NC in domain wall-based nanoelectronics.
\end{abstract}

\maketitle

\begin{figure*}[!t]
\center
\includegraphics [width=16cm] {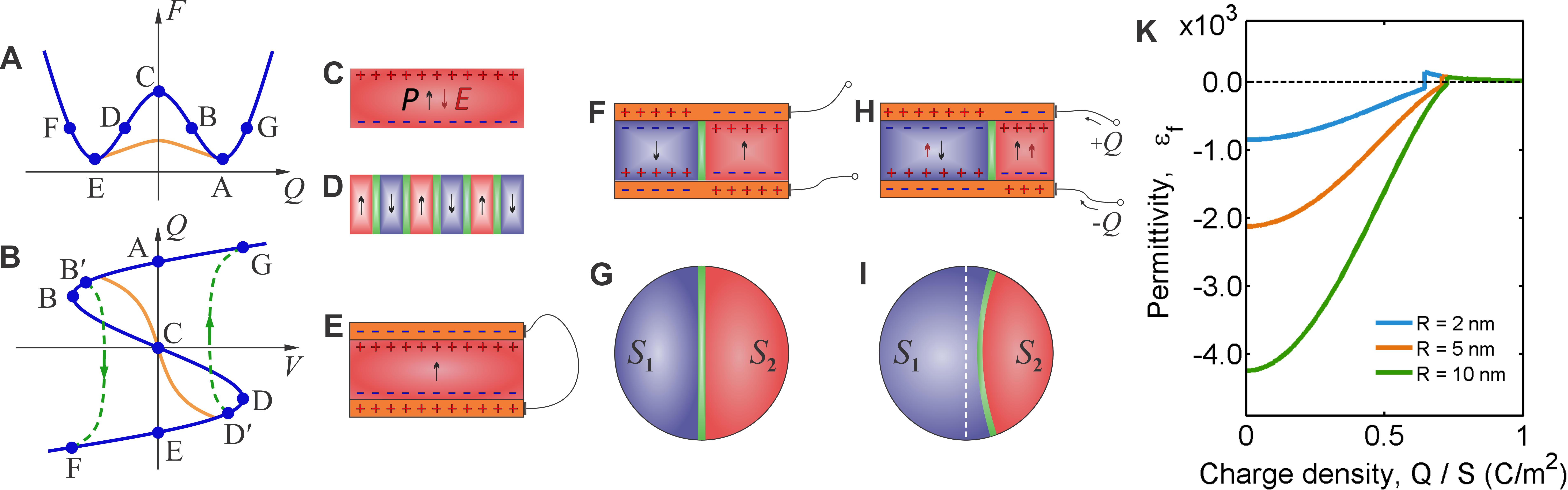}
\caption{Ferroelectric capacitor with negative capacitance. 
(\textbf{A})\,The plot of the Landau  free energy for the FE capacitor, $F$, as function of its charge, $Q$.
(\textbf{B})\,The corresponding charge-voltage characteristic, $Q$-$V$. The blue lines stand for the monodomain state, while the orange lines stand two-domain state with static NC. The dashed green curves depict the polarization switching process manifesting the  transient NC. 
(\textbf{C})\,The FE monodomain sample. Termination of polarization, $\mathbf{P}$, 
at the surface induces
surface depolarization charges and the depolarization field $\mathbf{E}$. 
(\textbf{D})\,The periodic domain structure. 
(\textbf{E})\,The FE sample with short-circuited electrodes, which recovers
the monodomain
structure. 
(\textbf{F}) Formation of two equal domains in a FE confined between disconnected electrodes redistributes the free charges and cancels the depolarization field.
(\textbf{G})\,Top view of the two-domain structure
with disconnected electrodes. 
(\textbf{H})\, Adding the charge, $Q$, pushes the DW towards the edge.
(\textbf{I})\,Top view of the FE capacitor with the displaced DW.
(\textbf{K})\,Negative differential permittivity ${\varepsilon} _{f}$ of the cylindrical FE PbTiO$_3$ nanodot capacitors of different radii as functions of the charge density, $Q/S$.  
The jumps in $\varepsilon_f$ from the negative to positive value mark the full polarization of the nanodot.}
\label{FigPrinciple}
\end{figure*}

 Over 40 years ago Rolf Landauer raised the question whether the
capacitance can be negative\,\cite{Landauer1976}, that is if the increase in the charge of the capacitor can decrease its voltage. Having stated that
ferroelectrics (FE) can harbor the
negative capacitance (NC) during the transient processes of switching, Landauer indicated that the very existence of the static NC as a part of steady state is challenged by 
instability against spontaneous domain formation. 
The idea of ferroelectric-based negative capacitance (NC) devices\,\cite{Salahuddin2008,Khan2015} brought back the suggestion of engineers of early 1930-s on the exploration of circuits with 
negative elements\,\cite{Verman1931,Behr1932,Terman1934} and 
ignited an explosive research activity, see\,\cite{Catalan2015,Ng2017,Hoffmann2018domains} for a review. 
The NC would naturally arise in the Landau description of the
uniformly polarized FE, where the free energy, $F$ as a function of
polarization, $P$, or capacitor charge, $Q$, assumes a double well form, see
Fig.\,\ref{FigPrinciple}a. A standard approach based on the minimization of $F$ at the given
voltage $V$ yields the dependence $Q(V)$, where the induced charge
represents the system's response to the applied voltage. The resulting
charge-voltage characteristic has the form of an S-shaped curve, GABCDEF,
shown in Fig.\,\ref{FigPrinciple}b. Being driven by the applied voltage, the capacitor
switches between the two stable branches GAB and DEF, corresponding to the
up- and down- oriented FE polarizations, in a hysteretic fashion. The
differential capacitance, $C=dQ/dV$, is positive when the system falls into 
either of these two polarized states. Remarkably, $C$ reverses its sign when
the system traverses along the intermediate branch BCD, having the negative
slope. However, since the BCD path of the $Q(V)$ dependence corresponds to
the concave segment in the energy profile (Fig.\,\ref{FigPrinciple}a), this intermediate
branch of the S-curve is unstable, hence the static NC regime is unreachable
for a voltage-driven capacitor.
In an experiment, the switching instability may occur also even before reaching the turning points B and D as a result of the reverse domain nucleation and accelerated
 growth. The corresponding time-dependent evolution of the charge and voltage drop would then follow the dynamical $Q(V)$ trajectory, depicted by B$^\prime$F and D$^\prime$G dashed segments in Fig.\,\ref{FigPrinciple}b, with the negative slope at the beginning of the switching\,\cite{Smith2017,Hoffmann2018domains}.
 It is this negative slope that is commonly viewed as a manifestation of NC, which, however, in this case, has a transient and irreversible character.

What can stabilize the static NC, is setting the charge, $Q$, as a driving
parameter controlling the state of the capacitor, so it is $V$ that becomes
the system's response. This would reverse the S-shaped $Q(V)$ dependence of the
monodomain FE state into an N-shaped $V(Q)$ function that maintains the
negative sign of $dQ/dV$ at low $Q$ but has no bistability anymore. Yet, however attractive, this approach may not to straightforwardly work since
the charge-driven system would likely split
into the multidomain state\,\cite{Landauer1976}.
For instance, the paraelectric uniform state C
at $Q=0$ will break up into the lower-energy FE states A and E comprising the domains with alternating oppositely oriented
polarization. 
This effect is often thought to be detrimental to steady state NC\,\cite{Cano2010,Hoffmann2018nanoscale}.

However the situation appears not hopeless once the delicate balance between the domain wall (DW) and electric field  energies is  taken into account. To see that, let us consider the mechanism of domain formation in a FE sample more in details. The spontaneous polarization, $P_{0}$, inside the monodomain FE flat
cell with uniaxial
anisotropy generates the depolarization surface charges, 
$+P_{0}$, at the top and $%
-P_{0}$ at the bottom of the cell, which induce the depolarization field $%
E_{dep}=-P_{0}/\varepsilon _{0}\varepsilon _{\parallel }$, directed opposite
to the polarization (Fig.\,\ref{FigPrinciple}c). Here $\varepsilon _{\parallel }$ is
the permittivity of the FE material along $P$, and $\varepsilon _{0}$ is the
absolute perimittivity of the vacuum. An additional energy associated with
this field disfavors the uniform polarization and splits the system into the
alternating sequence of the up/down polarized domains (Fig.\,\ref{FigPrinciple}d). 
Such a domain texture
was found in ferroelectric-dielectric (FE-DE) heterostructures\,\cite{Bratkovsky2000,Streiffer2002,Kornev2004,Lukyanchuk2005,Zubko2010}. It was demonstrated,\cite{Bratkovsky2001,Zubko2016,Lukyanchuk2018} that the FE layers harboring domains possess a negative permittivity, i.e. the average intrinsic
field there is directed opposite to the electrostatic displacement. 
However this negative permittivity effect as is remains hard to utilize for a NC device, since the metallic electrodes when placed on the surfaces of the FE layer would effectively screened the depolarization fields induced by domains and the domains themselves would disappear right off (Fig.\,\ref{FigPrinciple}e). 


Here we step into the breach and demonstrate that the problem of the implementation of the steady reversible NC is achieved by about 10\,nm
nanodot FE capacitor, endowed with the \textit{two-domain} ground state. 
Splitting the monodomain state into the two-domain ground state and the subsequent motion of the DW, results in the modified free energy curve and corresponding modification of the $Q(V)$ dependence shown by orange curves in Fig.\,\ref{FigPrinciple}a,b. These characteristics retains the features of the S-curve of the monodomain system, hence the steady state NC. 
We consider, for concreteness, the cylindrical FE nanodot of
radius $R$ and thickness $d_f$ sandwiched between the disconnected metallic electrodes. 
Let the charge at the electrodes of the FE capacitor be zero, $Q=0$, Fig.\,\ref{FigPrinciple}f.
Then the energy of the system is indeed minimized by splitting the system into
the two equal-size domains with equal surfaces $S_{1}=S_{2}=S/2$, with $%
S=\pi R^{2}$, and down/up directed polarizations, $-P_{0}$, 
and $+P_{0}$ (Fig.\,\ref{FigPrinciple}g). These oppositely oriented polarizations generate surface depolarization charges of the
corresponding densities $\mp P_{0}$ at the top surface of the FE cell and
the charges of the opposite signs at the bottom surface. This, in turn,
induces redistribution of the free charges in the electrodes to compensate
electric fields of the surface depolarization charges. Since the domains are
of the equal size, the electroneutrality of the electrodes, $Q=0$, is
preserved and the average polarization and the corresponding depolarization
charges are zero. As the price to pay for eliminating the depolarization
energy is the DW energy, the least energy cost is provided by
splitting the bulk of the FE capacitor into the two equal size domains with
the minimal area of the DW. Since the depolarization energy is proportional
to the volume of the system, whereas the DW energy cost is proportional to
its area, the splitting into domains provides the gain in energy and drives
the system into the stable state.

\begin{figure*}[!t]
\center
\includegraphics [width=16cm] {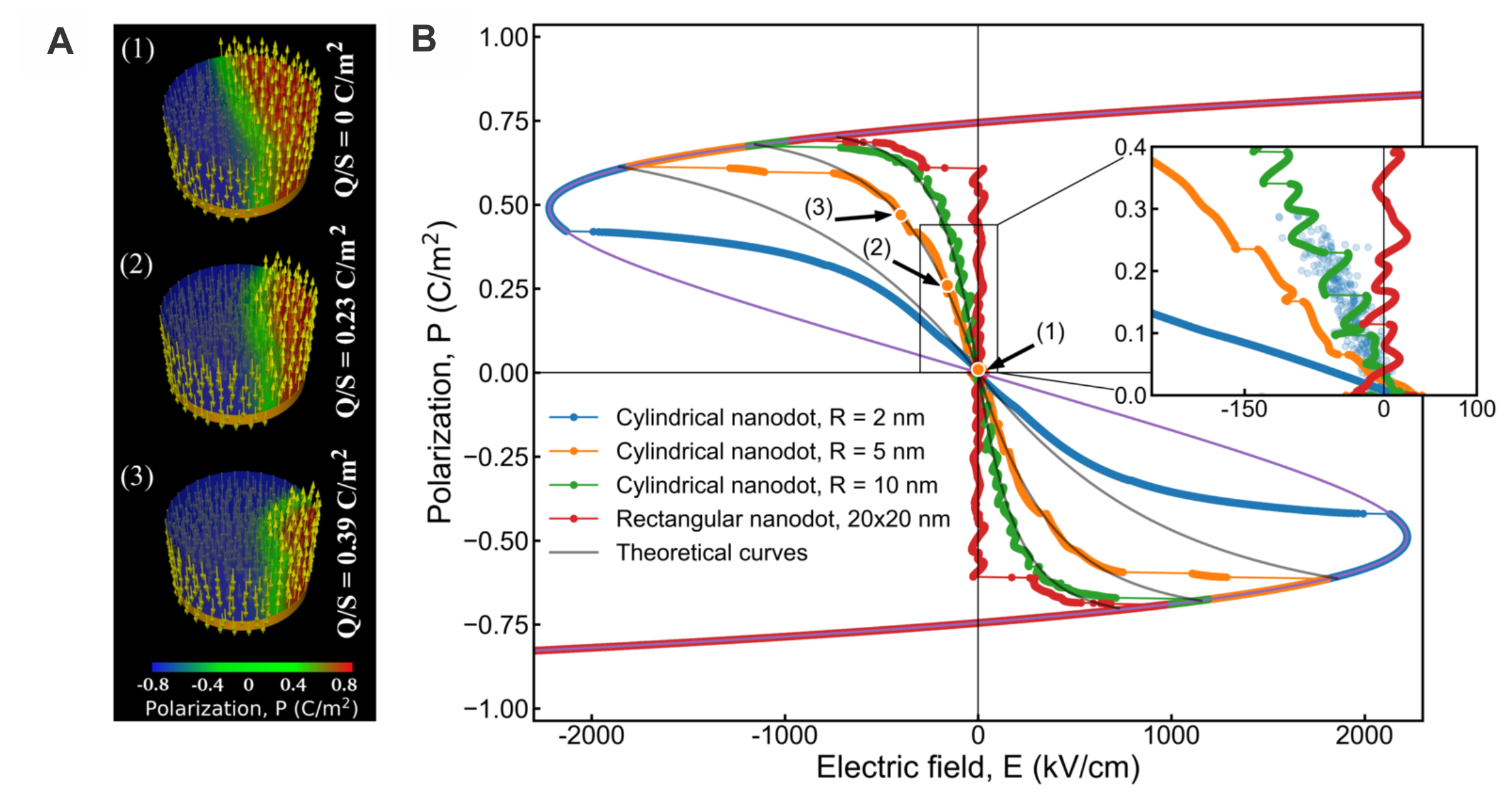}
\caption{ {Polarization $P$-$E$ characteristics for the FE nanodot capacitor. } 
(\textbf{A})\,The two-domain state in the
compressive-strained FE PbTiO$_3$ nanodot of radius $R=5$\,nm and thickness $%
d_f=5$\,nm as function of the electric charge, $Q$, on the electrodes is
obtained by the phase-field simulations. The configurations (1) -- (3)
correspond to points (1) -- (3) in the $P(E)$ plot shown in the panel (B). Arrows indicate directions of the polarization. The
orange rim depicts the bottom electrode, the top electrode is not shown. (1) Equal areas
at $Q=0$. (2) Displacement and bending of the DW at
finite $Q$. (3) On the brink of the complete disappearance of the
up-oriented domain. 
(\textbf{B})\,The $P$-$E$ characteristics
for 5\,nm thick nanodots of various lateral sizes and shapes are shown by
different colours. Theoretical curves are shown by solid lines. 
The details of the pinning-induced jerky motion of the DW  are displayed in the inset. 
For comparison, we present the results for quenching the nanodot from the
paraelectric phase (blue dots).
  }
\label{FigScurve}
\end{figure*}

Adding the charge on
electrodes, shifts the DW, which finds the
corresponding new equilibrium position (Fig.\,\ref{FigPrinciple}h).
To immediately see that the FE capacitor harboring the two-domain state has
a negative capacitance we note, that if one short-circuited the plates,
allowing for the free electrons to flow between the plates and charge the
capacitor to allow to compensate the uniform depolarization field, the
DW exits from the system which thus transits into the monodomain state with the lower energy Fig.\,\ref{FigPrinciple}e. This self-charging effect manifests the emergence of the NC. For the evaluation of the NC, let us place the small test charges, $+Q$ on the top
electrode and $-Q$ on the bottom electrode of the FE capacitor, Fig.\,\ref{FigPrinciple}h, (the electrodes
are disconnected) and follow its response. 
Charging the capacitor causes 
the displacement of the DW in such a way that the depolarization field arising due to the misbalance of the domain surfaces, $S_1$ and $S_2$, (Fig.\,\ref{FigPrinciple}i) would screen the electrical field induced by the test charges.  However, the DW overshoots the position of the exact field compensation and reverses the internal electrical field $E$ since the extra displacement of the DW decreases further its surface, hence the total surface energy, due to geometrical finiteness of the system. This is the physical origin of the NC.  
To carry out a quantitative analysis, we calculate the internal
 field, $E$, and find the integral capacitance as ${\tilde C}_{f}\equiv Q/V$ where $%
V=-d_{f}E$ is the potential difference between the top and bottom
electrodes; we use the tilde to distinguish ${\tilde C}_{f}$ from the differential capacitance ${C}_{f}\equiv dQ/dV$.

We obtain the integral FE capacitance in a standard form with the \textit{negative}, integral permittivity $\tilde{\varepsilon}_f$ (see Appendix): 
\begin{equation}
{\tilde C}_{f}=\varepsilon _{0}{\tilde \varepsilon} _{f}\frac{S}{d_{f}}\,,\;\;\;\;\;\;\;%
{\tilde\varepsilon}_{f}=-\frac{\pi }{4\nu }\psi (Q) \frac{R}{\xi _{0}}%
\varepsilon _{\parallel}\,.  \label{capacitance}
\end{equation}%
The integral permittivity,  $\tilde{\varepsilon}_f$, is a fundamental characteristic of the FE nanodot and appears also in the expression for the polarization curve, $P(E)=\varepsilon_0(\tilde{\varepsilon}_f-1)E$.  The dimensionless function $\psi(Q)$ reflects
the nonlinear charging effects, see Appendix. At small $Q\rightarrow 0$, one finds ${\tilde\varepsilon}_{f} \approx -\frac{8 }{3\pi \nu }\frac{R}{\xi _{0}}%
\varepsilon _{\parallel}$. 
The dimensionless parameter $\nu\simeq w_{DW}/w_{f}
$, with $w_{f}\simeq 2\xi _{0}P_{0}^{2}/\varepsilon _{0}\varepsilon
_{\parallel}$, characterizes the energy cost for the formation of the DW
normalized to the bulk energy of the FE stored in the volume
occupied by the DW. As a specific example, we consider the FE material PbTiO$_3$ with $\varepsilon
_{\parallel}\simeq 50$ and $\nu\simeq 0.1$, which is computed by the phase-field calculations, described in Methods. The DW width is taken of order of the FE coherence length $\xi_0\simeq 1$\,nm. 

Equation (\ref{capacitance})  that demonstrates the negative capacitance and permits to find the $Q$-$V$ and $P$-$E$ characteristics of the FE capacitor is the
central result of our work. It permits also to find the experimentally measured differential capacitance and the effective differential permittivity, related with the corresponding integral parameters via $C_f =\tilde{C}_f+V\left( d\tilde{C}_f/dV\right) $, $\varepsilon_f =\tilde{\varepsilon}_f+E\left( d\tilde{\varepsilon}_f/dE\right)$.
The behavior of the differential  permittivity,  as a function of the charge density, is shown in Fig.\,\ref{FigPrinciple}k for cylindrical nanodots of PbTiO$_3$ of radius $R=2$,\,5 and 10\,nm, respectively. 
The interval of the NC is extended over all range of the domain existence. 
The jumps in the $\varepsilon_f$ from the negative to the positive values correspond to the complete polarization of the FE capacitor to the monodomain state. 

\begin{figure*}[!t]
\center
\includegraphics [width=15cm] {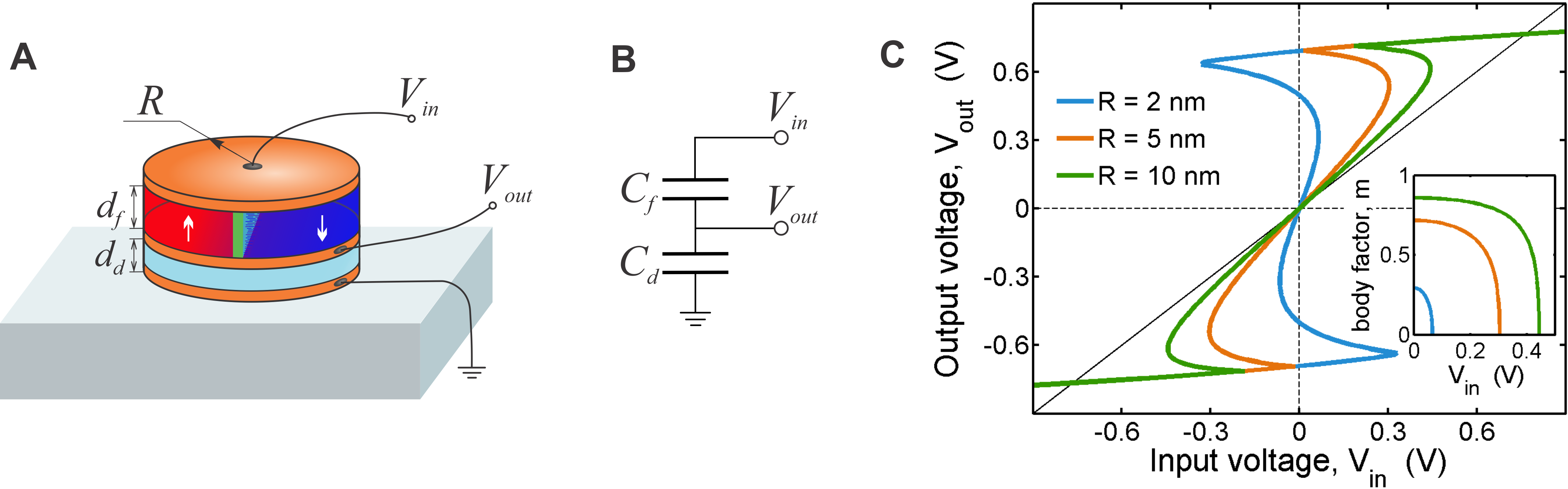} 
\caption{ {The NC-based amplifying circuit.} 
(\textbf{A})\, 
The device has a substrate-deposited Big Mac structure comprising the
 FE (carrying up- and down-polarization domains, shown by the red and dark blue, respectively and separated by the green DW), and the DE (light blue) nano-platelets. The FE and DE layers are confined between the three-parts
electrode shown in orange.
The top
electrode, strained uniaxial FE layer, and intermediate electrode
constitute the FE capacitor. 
The input voltage, $V_{in}$, is applied to the cover top electrode. 
The output voltage, $V_{out}$, is collected from the DE capacitor and exceeds the
input voltage, $V_{out}>V_{in}$. 
(\textbf{B})\,The equivalent scheme of the
circuit with $C_f<0$ corresponding to the FE and $C_d>0$ corresponding to the DE capacitors. 
(\textbf{C})\,The plots $V_{out}(V_{in})$ corresponding to different radii of PbTiO$_3$ nanodots. 
The inset demonstrates the differential body factors, $m=(dV_{out}/dV_{in})^{-1}$,  plotted as a function of $V_{in}$.}
\label{FigDevice}
\end{figure*}

We carry out the phase field modelling of the FE capacitor, consisting of
the cylindrical FE nanodot of PbTiO$_3$ sandwiched between two thin metallic
electrodes and uniaxially-strained by the SrTiO$_3$ substrate. The approach
rests on the relaxation minimization of the strain-renormalized
Landau-Devonshire functional, see Methods. The response of the 5\,nm
thick cylindrical FE capacitor of radius $R=5$\,nm with the two-domain
structure to the variations in charge $Q$ placed on the electrodes, is shown
in Fig.\,\ref{FigScurve}a. In compliance with the model considerations described above,
the DW departs from its bisector position at $Q=0$ (state (1)),
traversing the sample with simultaneous bending (state (2)). Finally the DW
exits the sample (state (3)), leaving behind the uniformly-polarized
monodomain state.
The corresponding $P$-$E$ characteristics of the FE
capacitors of different sizes and shapes are shown in the Fig.\,\ref{FigScurve}b.
Importantly, the negative slope of the $P(E)$ dependencies, that manifest the NC,  maintains well beyond the linear regime at $E\sim 0$ 
and holds during the entire process of the motion of the DW. Therefore, the NC extends over the entire electric field interval from zero to the coercive threshold  where the DW
leaves the sample and the $P$-$E$ characteristics turn around into the standard
 behavior of the monodomain sample. At variance, the multidomain-state nucleation switching process\,\cite{Smith2017,Hoffmann2018domains}, resulting in the transient differential NC as well as in the charge-controlled single nano-domain formation\,\cite{Sluka2017}, resulting in the static differential NC, occur in the intermediate vicinity of the coercive field and far from the zero-field steady-state, see Fig.\,\ref{FigPrinciple}b. 
 Inset in Fig.\,\ref{FigScurve}b demonstrates the jerky character of the 
 dynamics of the, DW due to pinning effects.
 
For cylindrical nanodots, the
negative slope of $P(E)$ decreases with the decreasing nanodot's radius.
The theoretical dependencies $P(E)$, calculated from Eq.\,(\ref{capacitance}), are shown by solid lines. They perfectly describe the results of simulations except for the smallest 2\,nm sample where the DW width is of order of the sample size. Similarly, it is the finite size of the DW that results in the slight deviation of the experimental behaviors from the theoretical predictions at the moment of the exit when the with of the DW compares to the size of the already nearly disappearing residual domain.

For the 
rectangular nanodot, the nearly infinite slope of $P(E)$ reflects that the energy
of the parallel-displaced straight DW almost does not depend on its
position. The jump to the more shallow negative-slope dependence $P(E)$ in
the pre-critical region corresponds to the abrupt re-orientation of the DW
from the parallel to the edge- to the corner,-enclosing configuration (see animation).

Having found the $Q$-$V$ characteristics of the FE capacitor we are now equipped to include it into  an amplifying NC circuit. 
A sketch of the proposed device is shown in Fig.\,\ref{FigDevice}a. The capacitor consists of DE and FE cylindrical nanodots, consequently grown on the insulating substrate and partitioned by the platelet metallic electrodes. The equivalent circuit (Fig.\,\ref{FigDevice}b), which is a customary proposed model implementation of the NC low-dissipation field effect transistor\,\cite{Salahuddin2008,Cano2010}, comprises the  two in-series capacitors, FE and DE with capacitancies $\tilde{C}_f$ and $C_d$, respectively. The operating voltage, $V_{in}$, applied to the top gate electrode, controls the state of the system, and in particular, the output voltage drop on the DE layer, $V_{out}$, is collected from the middle electrode. The bottom electrode is grounded. Once $\tilde{C}_f$ is negative and $C_d$ is positive, then applying the input voltage, $V_{in}$,  results in an amplification of $V_{out}$ given by 
$V_{out}=\tilde{m}^{-1}V_{in}$ where $\tilde{m}=1+C_{d}/\tilde{C}_{f}<1$ is the so-called body factor\,\cite{Salahuddin2008} and  $\tilde{C}_f(Q)$ is given by Eq.\,(\ref{capacitance}) with $Q=C_d V_{out}$.

Figure \,\ref{FigDevice}c shows, how the output voltage $V_{out}$ depends on the driving voltage $V_{in}$  for  PbTiO$_3$/SrTiO$_3$ cylindrical nanodots of radii $R=$2, 5 and 10\,nm, in which the DE layer is about half thin compared to the FE one, and $C_{d}=\varepsilon _{0}\varepsilon _{d}S/d_{d}$ with $\varepsilon_d \simeq 300$ for SrTiO$_3$. 
The working range of the device is confined between the negative and positive reverse points in the middle branch of $V_{out}(V_{in})$ curves where the two-domain state is locally stable. The reversal segments correspond to the unstable situation when the DW comes out from the sample. The upper and lower branches of the curves reflect the monodomain states with the positive differential capacity. As the working part of the $V_{out}(V_{in})$ dependence corresponding to the two-domain state lies above the $V_{out}=V_{in}$ bisectrix, the device exhibits the amplification effect, which, in the linear approximation, is given by the body factor 
\begin{equation}
\tilde{m}=1-\frac{3\pi\nu }{8 }\frac{\varepsilon _{d}}{%
\varepsilon _{\parallel }}\frac{d_{f}}{d_{d}}\frac{\xi _{0}}{R}\,. \label{bodyexpr}
\end{equation}
For nanodots with $R=5$\,nm the estimate gives $\tilde{m}\simeq 0.69$, which can be improved by further material and geometry optimization, for instance, via selecting the proper ferroelectric material with the low permittivity and high DW energy (expressed through the factor $\nu$), while the DE layer is to be done as thin as possible and should have the highest permittivity available. At the same time, the nanodot radius should be chosen as small as possible, bearing in mind the intertwined relationships between the parameters (that e.g. $2R>d_f, \xi_0$ should hold). Further improvement can be achieved by going into the nonlinear regimes near instability reversal points in the $V_{out}(V_{in})$ dependencies, i.e. to the parts corresponding to the flattering of $P(E)$ curves in Fig.\,\ref{FigScurve}b. There, the respective values of the differential body factor, $m=(dV_{out}/dV_{in})^{-1}$, become appreciably smaller, as shown in the inset to Fig.\,\ref{FigDevice}c.


\section*{Appendix} 

\subsection*{Phase-field calculations}

 The description of the uniaxially-strained perovskite FE platelet, rests on the
minimization of the functional,
\begin{equation}
F=\int \left[ f_{GL}(T,u_{m},P_{i})+f_{grad}(\partial _{i}P_{j})-\frac{1}{2}%
P_{z}E\right] d^{3}r,
\label{LDF}
\tag{S1}
\end{equation}
where $i,j=x,y,z$, the uniform electric field is produced by the electrode- and depolarization surface charges, $E=-\left( Q/S+\overline{P}_{z}\right) /\varepsilon _{0}\varepsilon_{i}$, and the bar denotes averaging of the polarization at the platelet surface.  The uniform Ginzburg-Landau part of the functional, $f_{GL}$, is taken in the strain-renormalized form with the appropriate for the PbTiO$_3$ material coefficients\,\cite{Pertsev1998} and the compressive strain $u_m\simeq-0.013$ is produced by the SrTiO$_3$ substrate. The gradient part of the functional, $f_{grad}$, is taken from\,\cite{Li2001} with the PbTiO$_3$-specific coefficients.   

Numerical phase-field calculations were performed using the 
FERRET package\,\cite{Mangeri2017} for the open-source multiphysics object-oriented 
finite-element framework MOOSE\,\cite{Gaston2009} by solving time-dependent 
 relaxation equation $-\gamma \, \partial P_i/\partial t=\delta F / \delta P_i$ with free boundary conditions for $P_i$ at the sample surface, coupled with electrostatic equations. 
 Random small distribution of $P_i$ was chosen
as initial condition.

\subsection*{Domain wall geometry}

\begin{figure*}[!t]
\center

\includegraphics [width=15cm] {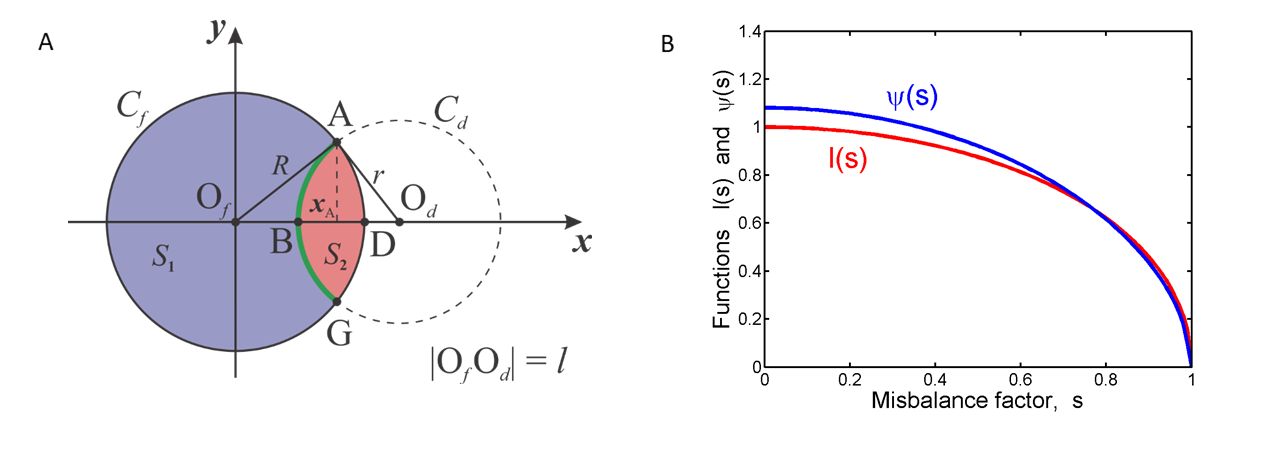} 
\caption{ (\textbf{A})\,The geometry of the system. (\textbf{B})\,The functions $l(s)$ and $\psi(s)$  where $s=(S_1-S_2)/S$ is the misalliance parameter. }
\label{FigSuppl}
\end{figure*}

In this section, we find the dependence of the DW length $L$ on the
misbalance factor $s=\left( S_{1}-S_{2}\right) /S$, where $S_{1\text{ }}$and 
$S_{2}$ are the surfaces of domains with down- and up-oriented polarization,
respectively, and $S=S_{1}+S_{2}=\pi R^{2}$ is the total cross-section area
of the FE capacitor, $R$ is the nanodot radius. The sketch of the cross-section plane $xy$
(Fig.\thinspace {\color{red} S\ref{FigSuppl}}A) presents the geometry of the system. The DW
has a cylindrical-arc shape minimizing the DW surface tension energy at
fixed $s$. The arc makes the right contact angle with the nanodot surface to
eliminate the tangential drag caused by the surface tension. 
  The bended DW can
be seen as the segment ABG obtained by the crossing of the FE nanodot circular
boundary, $C_{f}$, with an external circle $C_{d}$ of radius $r$,
both circles being mutually orthogonal. Selecting the origin at the center of 
$C_{d}$ and axis $x$ and $y$ along and perpendicular to the line, connecting
the centers O$_f$ and O$_d$ of the circles $C_f$ and $C_{d}$ respectively,  we write the geometrical equations for
these circles  as 
\begin{gather}
y_{f}^{2}+x_{f}^{2} =R^{2}, \label{circles} \tag{S2} \\
y_{d}^{2}+(x_{d}-l)^{2} =r^{2}.  \notag
\end{gather}
Here $l=|O_fO_d|=\sqrt{R^{2}+r^{2}}$ is the distance between the centers of the
circles. A simple geometrical consideration shows that the length of the DW
can be expressed as $L=2R\,l(\rho )$ using the dimensionless function: 
\begin{equation}
l(\rho )=\rho ^{-1}\arctan \rho ,  \label{L}
\tag{S3}
\end{equation}%
of the dimmensinless parameter $\rho =R/r$. The surface $S_{2}$, is
calculated as area confined by the segments ABG and ADG  as $%
S_{2}=2\int_{x_{B}}^{x_{A}}y_{f}(x_{f})dx_{f}+2%
\int_{x_{A}}^{x_{D}}y_{d}(x_{d})dx_{d}$, where the dependencies $%
y_{f}(x_{f}) $ and $y_{d}(x_{d})$ are given by Eqs.$\,$(\ref{circles}) and $%
x_{B}=l-r$, $x_{A}=R^{2}/l$ and $x_{D}=R$. Taking $S_{1}=S-S_{2}$ we obtain: 
\begin{equation}
s=\frac{2}{\pi }\left[ \rho ^{-1}+\left( 1-\rho ^{-2}\right) \arctan \rho %
\right] .  \label{s}
\tag{S4}
\end{equation}
Equations (\ref{L}), (\ref{s}) provide the dependence $L(s)$ parametericaly
where the running parameter $\rho $ changes from from $0$, which corresponds
to the straight bisectoral DW at $s=0$, to $\infty $, which corresponds to
the small semi-circle DW, leaving the sample at $s=1$. The graphic
presentation of $l(s)$ is shown in Fig.\,{\color{red} S\ref{FigSuppl}}B. The behaviour of $l(s)$ at small 
$s$ is found by the proper expansions of expressions (\ref{L}) and (\ref{s}): 
\begin{equation}
l(s)\simeq 1-\frac{3\pi ^{2}}{64}s^{2}.  \label{limits}
\tag{S5}
\end{equation}

\subsection*{Distribution of charge and field in FE capacitor}

The depolarization charges at the surface of FE cell have the densities $%
\sigma _{1}^{dep}=-P_{0}$ and $\sigma _{2}^{dep}=+P_{0}$ for the domains
with the down- and up-oriented polarization, $-P_{0}$ and $+P_{0}$,
respectively (we consider here the upper surface for definiteness). The test
charge, placed at the electrodes is redistributed between areas $S_{1}$ and $%
S_{2}$ with the densities $\sigma _{1}^{el}$ and $\sigma _{2}^{el}$, keeping
the value of the test charge, $Q$, constant, 
\begin{equation}
Q=\sigma _{1}^{el}S_{1}+\sigma _{2}^{el}S_{2}.  \label{Qconserv}
\tag{S6}
\end{equation}%
This distribution should ensure the equality of the electrostatic potential over
the whole electrode area. The latter condition is equivalent to the uniformity of the distribution of the total charge, $q$, that includes both
the depolarization and free charges and to the
uniformity of the distribution of the field  inside the FE
capacitor.  Accordingly, the density of the total
charge is equal at the surfaces $S_{1}$ and $S_{2}$ and is given by 
\begin{equation}
\sigma ^{tot}=q/S=\sigma _{1}^{el}-P_{0}=\sigma _{2}^{el}+P_{0}.
\tag{S7}
\label{sigmatot}
\end{equation}%
Equations  (\ref{Qconserv}) and (\ref{sigmatot}) allow to
calculate $q$, 
\begin{equation}
q=Q-sSP_{0},  \label{q}
\tag{S8}
\end{equation}%
where the depolarization charge entering with the weight $s$ that accounts for the
alternation of the polarization orientation in domains. Finally we find the internal field in FE capacitor, 
\begin{equation}
E(s)=-\frac{\sigma ^{tot}}{\varepsilon _{0}\varepsilon _{\parallel }}=-\frac{%
q}{S\varepsilon _{0}\varepsilon _{\parallel }}=-\frac{1}{\varepsilon
_{0}\varepsilon _{\parallel }}\left( Q/S-sP_{0}\right)   \label{field}
\tag{S9}
\end{equation}%
as function of the factor $s$.

\subsection*{Calculation of the capacitance}
The energy of the system reads: 
\begin{equation}
\mathcal{E}=\frac{V_{f}}{2}\varepsilon _{0}\varepsilon _{\parallel
}E^{2}+d_{f}L\,w_{DW}\,,  \label{energy}
\tag{S10}
\end{equation}%
where the first term describes the electrostatic energy contribution, while
the second term represents the energy of the DW. Here  $V_{f}=\pi R^{2}d_{f}$ is the FE capacitor volume, $w_{DW}=2\xi _{0}\nu P_{0}^{2}/\varepsilon
_{0}\varepsilon _{\parallel }$ is the surface tension of DW, $L$ is the DW length, and $d_{f}L$ is
the DW\ area. Both, the field, $E$, and the domain length, $L$, are
parametrized via the misbalance factor $s$. The dependence $%
E(s)$ is linear and is given by Eq.\,(\ref{field}). The DW length is expressed via the shown in Fig.\,{\color{red} S\ref{FigSuppl}}B function $l(s)$ 
as $L(s)=2Rl(s)$. 

Considering $s$ as a variational parameter and minimizing $\mathcal{E}$ we obtain the equation%
\begin{equation}
s+\frac{4\nu }{\pi }\frac{\xi _{0}}{R}l^{\prime }(s)=\frac{Q}{SP_{0}},
\tag{S11}
\label{itter}
\end{equation}%
that can be solved by recurrence, since the second term in the right-hand
side is smaller than the first one. (It tends to zero when the DW energy
associated with $\nu $ vanishes).

After decomposition $s=s_{0}+s_{1}$, with $s_{0}=Q/SP_{0}$ and $\left\vert
s_{1}\right\vert \ll \left\vert s_{0}\right\vert $ we obtain 
\begin{equation}
s=\frac{Q}{SP_{0}}-\frac{4\nu }{\pi }\frac{\xi _{0}}{R}l^{\prime }(s_{0})=%
\frac{Q}{SP_{0}}\left( 1+\frac{4\nu }{\pi \psi (s_{0})}\frac{\xi _{0}}{R}%
\right) 
\tag{S12}
\end{equation}%
where the function $\psi (s)$ is defined from $l(s)$ as $\psi (s)\equiv
-s/l^{\prime }(s)$. Its graphical presentation is given at Fig.\,{\color{red} S\ref{FigSuppl}}B. At small 
$s$ $\psi (s)\simeq \frac{32}{3\pi ^{2}}\simeq 1.08$ as it follows from
Eq.\thinspace (\ref{limits}).

From (\ref{field}) we obtain the internal field
\begin{equation}
E=\frac{1}{\varepsilon _{0}\varepsilon _{\parallel }}\frac{Q}{S}\frac{4\nu }{%
\pi \psi (s_{0})}\frac{\xi _{0}}{R}  \label{intfield}
\tag{S13}
\end{equation}
and, finally%
\begin{equation}
{\tilde C}_{f}=\frac{Q}{V}=\frac{Q}{-d_{f}E}=-\varepsilon
_{0}\varepsilon _{\parallel }\frac{S}{d_{f}}\frac{\pi }{4\nu } \psi (s_{0})
\frac{R}{\xi _{0}}  \label{capas}
\tag{S14}
\end{equation}
\section*{Acknowledgments}
 This work was supported by H2020-RISE-ENGIMA 
 action (I.L. A.S. and A.R.), by internal grant № VnGr-07/2017-23 of Southern Federal University, Russia (A.R. and Y.T.) and by the U.S. Department of Energy, Office of Science, Basic Energy Sciences,
Materials Sciences and Engineering Division (V.M.V. and partially I.L.).

\end{document}